\documentclass[journal,twocolumn,10pt]{IEEEtran}

\usepackage{cite}
\usepackage[pdftex]{graphicx}
\usepackage{amsmath}
\usepackage{amssymb}
\usepackage{mathtools}
\usepackage[caption=false,font=footnotesize]{subfig}
\usepackage{url}
\usepackage{enumitem}

\usepackage{xspace}
\usepackage{xcolor}
\usepackage[acronym,nogroupskip,nonumberlist,nopostdot]{glossaries}
\makeglossaries

\newacronym{ai}{AI}{artificial intelligence}
\newacronym{cli}{CLI}{command line interface}
\newacronym{csi}{CSI}{channel state information}
\newacronym{isac}{ISAC}{integrated sensing and communication}
\newacronym{ism}{ISM}{industrial, scientific, and medical}
\newacronym{lan}{LAN}{local area network}
\newacronym{mcp}{MCP}{model context protocol}
\newacronym{ml}{ML}{machine learning}
\newacronym{oai}{OAI}{OpenAirInterface}
\newacronym{ofdm}{OFDM}{orthogonal frequency-division multiplexing}
\newacronym{ota}{OTA}{over-the-air}
\newacronym{otfs}{OTFS}{orthogonal time frequency space}
\newacronym{pawr}{PAWR}{Platforms for Advanced Wireless Research}
\newacronym{poe}{PoE}{power over Ethernet}
\newacronym{rpc}{RPC}{remote procedure call}
\newacronym{sdr}{SDR}{software-defined radio}
\newacronym{vpn}{VPN}{virtual private network}

\newcommand{\sdr}{\gls{sdr}\xspace}
\newcommand{\sdrs}{\glspl{sdr}\xspace}
\newcommand{\ml}{\gls{ml}\xspace}
\newcommand{\ai}{\gls{ai}\xspace}
\newcommand{\vpn}{\gls{vpn}\xspace}
\newcommand{\cli}{\gls{cli}\xspace}

\newcommand{\figref}[1]{Fig.~\ref{#1}}

\NewDocumentCommand{\rev}{+m}{#1} 

\begin{document}

\title{\rev{RemoteRF: A Global, Open-Source Sandbox\\for Wireless Research and Education}}

\author{Ethan Y.~Ge and Ian P.~Roberts\\University of California, Los Angeles\\\texttt{www.remoterf.net}}

\maketitle

\begin{abstract}

\Glspl{sdr} are powerful tools for research and education in wireless communications, but their cost and complexity put them out of reach for many researchers and institutions worldwide.
To address this, we introduce RemoteRF, an open-source platform for creating large-scale testbeds of distributed \sdrs{} \rev{and multi-modal peripherals} that can be managed by a single server and shared across many users.
\rev{Through a reservation-based framework, users can control devices over the internet via Python scripting, with only device-related commands routed to the server.
In research, RemoteRF can be used to evaluate and reproducibly benchmark new physical-layer and full-stack 5G/6G techniques, collect real-world data for machine learning, and give AI agents controlled access to hardware for autonomous experimentation.}
In education, RemoteRF allows students in virtually any sized class to share a handful of \sdrs to learn communication fundamentals through hands-on implementation with actual radios.
The software powering RemoteRF is open-source and extensively documented online, allowing anyone to deploy their own instance in a matter of minutes.
\rev{Beyond local use, instances can also connect to the RemoteRF Global network to share hardware across partner institutions or make devices publicly accessible worldwide.}
Since being deployed at UCLA, RemoteRF has logged over 4,000 hours of use across more than 250 researchers and students.
\end{abstract}

\begin{IEEEkeywords}
Software-defined radios, platforms, testbeds, prototyping, measurements, communications, sensing, education.%
\end{IEEEkeywords}

\glsresetall

\section{Introduction} \label{sec:introduction}

For decades, mathematical modeling and simulation have been the primary tools used by countless researchers to invent new wireless communication techniques.
Similarly, engineering students have long been taught the fundamentals of wireless communication through blackboard lectures and problem-solving built on idealized theory.
While this approach to both research and training has yielded the incredible wireless networks we have today, fundamental progress is slowing by several measures, raising questions about where future wireless breakthroughs will come from~\cite{bjornson_2026_what_remains_human,dohler_2011_phy_dead}.
Although \gls{ml} and \gls{ai} have sparked new momentum, these methods are still predominantly trained and tested on synthetic data generated by the same idealized models and simulators~\cite{marti_2026_sionna_bench}.

While sound models and simulation tools will forever play important roles in wireless, incorporating real hardware platforms into research and education stands to revolutionize the field~\cite{bjornson_2026_what_remains_human}.
\rev{\Glspl{sdr} and multi-modal peripherals like radars, lidars, and cameras, for instance, can be used in research to take measurements, experimentally evaluate new techniques, and collect real-world data to train and test \ml/\ai models~\cite{oshea_2018_ota}.
Hardware access will also be critical to enabling future \ai-driven research, where agents autonomously run experiments and refine algorithms on real-world wireless systems~\cite{bjornson_2026_what_remains_human}.}
In education, \sdrs can be used by students to complete {active learning} lab exercises by implementing fundamental concepts on actual radios, preparing them for careers where they are tasked with translating ideas from theory to practice \cite{rondeau_2016_sdr}.

\subsection*{Key Barriers to Entry: Cost, Complexity, and Coordination}
Despite the proven value of \sdrs in research and education~\cite{rondeau_2016_sdr,collins_2018_sdr4engineers}, the vast majority of wireless researchers and engineering students across the globe still lack access to them, largely due to cost.
A single \sdr transceiver can cost anywhere from a few hundred to tens of thousands of US dollars (USD), putting even the most affordable models out of reach for many, especially when multiple devices are needed for large research testbeds or classes.
Beyond cost, setting up the necessary drivers/dependencies and writing code to control \sdrs introduces additional hurdles, particularly for first-time users.
This worsens when conducting experiments that require coordination across multiple \sdrs at once, especially in large-scale testbeds where devices are geographically distributed.
\rev{Sharing a limited pool of devices across many users compounds these challenges, as it requires a way to schedule access fairly, manage user permissions, keep the underlying systems secure, and ensure spectrum regulations are met.}
Altogether, these financial and logistical challenges present substantial barriers to incorporating \sdrs into research or education, stifling progress in wireless.

\begin{figure*}[!t]
    \centering

    \includegraphics[width=\textwidth]{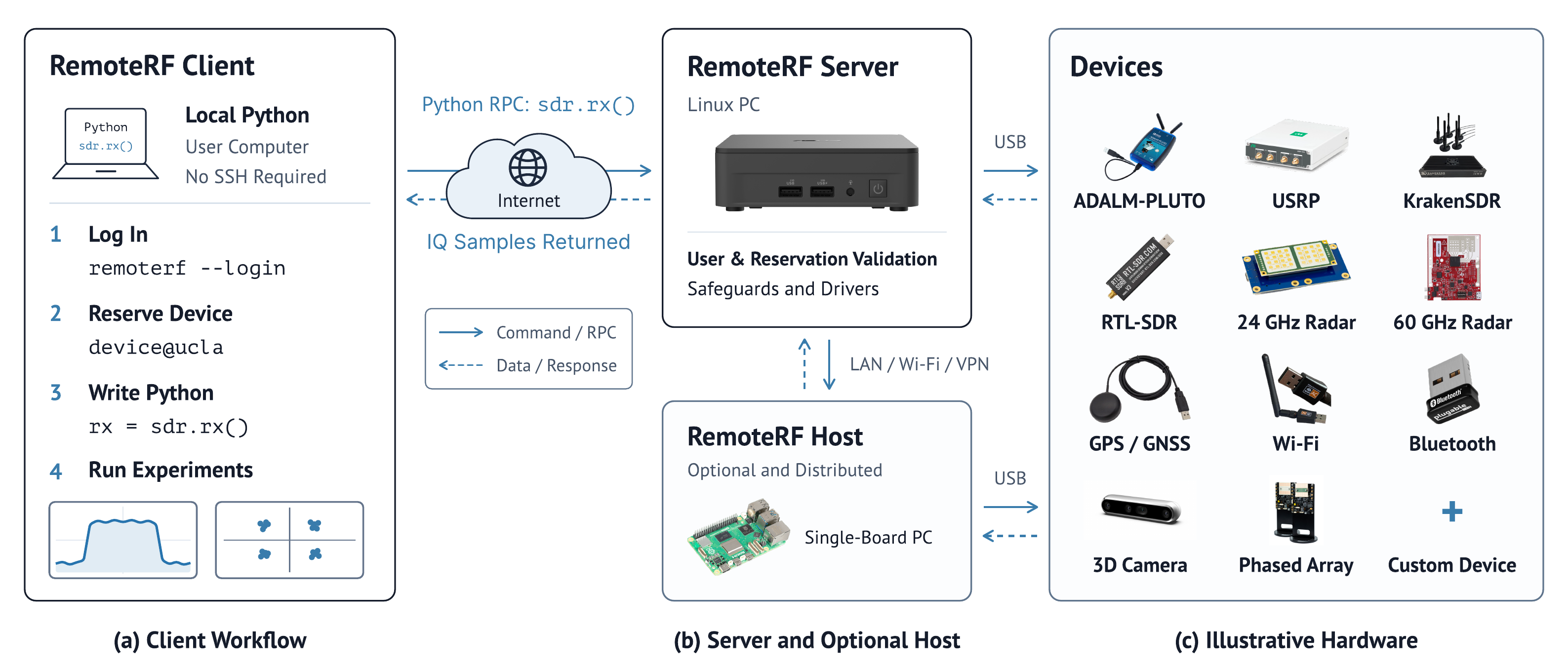}
    \caption{\rev{Overview of the RemoteRF platform: (a)~four-step client workflow, (b)~server and optional host executing device commands via remote procedure calls (RPCs), and (c)~example devices connected to the server or host.}}
    \label{fig:remoterf_system_overview}
\end{figure*}

\subsection*{Where Existing Solutions Fall Short}
Several efforts have sought to address this problem through the creation of shared platforms that allow users to \textit{remotely} access \sdrs over the internet.
Among the most popular are large-scale testbeds supported through the US National Science Foundation's \gls{pawr} program~\cite{pawr}---such as POWDER~\cite{breen2020powder}, COSMOS~\cite{raychaudhuri2020challenge}, and AERPAW~\cite{marojevic2020advanced}---which have been created primarily for research use, not education.
\rev{These platforms require prior approval to access and rely on dedicated infrastructure that would be challenging for an individual lab or instructor to replicate.
Meanwhile, smaller university testbeds such as the ORBIT radio grid at Rutgers~\cite{raychaudhuri2005overview} and the Arena testbed at Northeastern~\cite{bertizzolo2019arena}, along with educational platforms like WebLab-Deusto~\cite{orduna2018weblab} and the RHLab remote \sdr laboratory~\cite{inonan2023rhlab}, are typically custom, one-off setups that are not packaged or documented for others to deploy.
Furthermore, existing platforms all rely on remote-execution workflows in which users develop and run code within the testbed's own infrastructure, accessing allocated resources over SSH or through provisioned virtual machines (VMs).
Executing code on the platform itself slows iteration and shapes the user experience around the platform's constraints rather than their own local tools.
Finally, all of these platforms operate as standalone silos, with no simple way to share hardware across partner institutions or the general public.}

\begin{figure*}[!t]
	\centering
	\includegraphics[width=\textwidth]{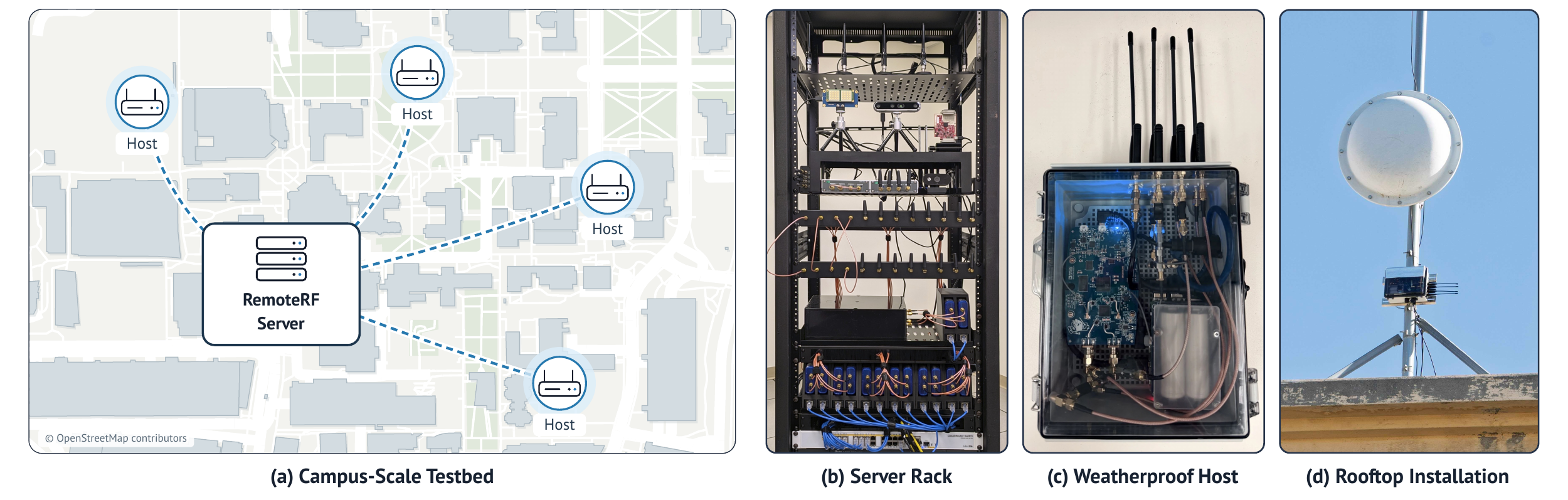}
	\caption{\rev{Campus-wide testbed architecture and hardware on the UCLA deployment: (a)~a central RemoteRF server managing distributed hosts across campus, (b)~the central server rack, (c)~a weatherproof outdoor host enclosure, and (d)~a rooftop installation.}}
	\label{fig:edgehosts}
\end{figure*}

\section{The RemoteRF Platform}
This article introduces the RemoteRF platform, which aims to dramatically lower the barrier to entry for those wishing to incorporate \sdrs{} \rev{and multi-modal peripherals} into their research or instruction, especially at scale.
RemoteRF accomplishes this by providing a systematic way to centrally manage and remotely access devices over a network connection.
This allows researchers or university instructors to create large-scale testbeds of multiple devices that can be seamlessly shared across many users, such as members of a research group or students in an engineering course.
\rev{The principal advantages of RemoteRF over existing alternatives are its low cost, simplicity, open-source software, local code execution, dual use in research and education, and support for cross-institutional and public device sharing.}

\subsection*{Overview of RemoteRF}
The RemoteRF platform is comprised of four distinct components, each of which is shown in \figref{fig:remoterf_system_overview}.
\begin{itemize}
    \item \rev{\textbf{Device:} A RemoteRF device is an \sdr, multi-modal peripheral, or virtually any instrument (or even software process) that can be controlled through Python.}
    \item \textbf{Server:} The RemoteRF server is a Linux computer with one or more connected devices, typically deployed on a university campus or in a research lab.
    \item \textbf{Client:} A RemoteRF client connects to the RemoteRF server and enables a user to remotely access the server's devices through Python via a network connection.
    \item \textbf{Host:} A RemoteRF host is an optional component that links one or more devices to the server over Wi-Fi, \gls{lan}, or \vpn{} instead of a direct wired connection like USB.
    A host can be a lightweight single-board computer such as a Raspberry Pi, making it easy to distribute devices across a room, building, or campus, as illustrated in \figref{fig:edgehosts}.
\end{itemize}

A typical RemoteRF deployment consists of one server, multiple devices, many clients, and optionally one or more hosts.
\rev{Setting up a RemoteRF server amounts to connecting the desired devices---either directly (e.g., over USB) or remotely via hosts---and installing the server software via \texttt{pip}:
\begin{center}
\vspace{-2pt}
\texttt{pip install remoterf-server}
\vspace{-2pt}
\end{center}
\noindent RemoteRF client software is installed on users' personal machines in the same way:
\begin{center}
\vspace{-2pt}
\texttt{pip install remoterf}
\vspace{-2pt}
\end{center}
\noindent This provides users with a \cli to connect to the server and a Python library to interface with its devices.
After registering an account through the \cli, a typical user workflow consists of four simple steps, as shown in \figref{fig:remoterf_system_overview}a: log in, reserve a device, write a Python script, and run experiments.}

\subsection*{Preserving Local Experimentation}
A key design goal of RemoteRF was to make remote \sdr use feel as close as possible to \textit{local} experimentation, allowing users to control networked \sdrs through Python scripts that closely resemble those written for directly connected hardware.
When users run these Python scripts on their personal computers, most of the code executes locally, but any commands that involve devices are routed over the network to the RemoteRF server as \glspl{rpc} for execution; any return values are passed back to the user for processing, plotting, and analysis.
In stark contrast to existing platforms~\cite{breen2020powder,raychaudhuri2020challenge,marojevic2020advanced,raychaudhuri2005overview,bertizzolo2019arena,orduna2018weblab,inonan2023rhlab}, this allows users to develop, debug, and iterate rapidly on their own personal computers using familiar tools, workflows, and compute resources (such as GPUs), rather than within a VM or over SSH.
\rev{At the same time, routing device interactions strictly through \glspl{rpc} streamlines device access and isolates the server from user code: clients require no vendor-specific hardware drivers, never receive shell access to the server or hosts, and can only invoke explicitly exposed methods gated by an active reservation, user-group permissions, and server-side lockouts.}
This separation also has the added benefit of reducing the computational load on the server, allowing it to accommodate more users and devices.

\subsection*{Reservation-Based Access}
A central component of the RemoteRF platform is its reservation framework, which enables many users to systematically share limited hardware and schedule experiments in advance.
Rather than accessing devices on a first-come, first-served basis, users reserve specific devices for desired time slots through the RemoteRF \cli.
A reservation grants the user exclusive access to a device during that window, allowing them to reference it in Python \rev{either by its human-readable handle (such as \texttt{usrp@ucla})} or by a unique reservation token.
To ensure fairness across users, administrators can also limit the duration and number of reservations each user may hold.

\subsection*{User Groups and Enrollment Codes}
Given that RemoteRF is a shared platform, it is natural for an administrator to want to provide different levels of access for different groups of users.
\textit{User groups} can serve such a purpose, where each group can be given unique permissions and privileges, such as access to certain devices and custom reservation limits.
Users can join a user group by entering an \textit{enrollment code}---either when creating their account or at any time afterward through the \cli---which automatically grants them that group's permissions.
This allows an administrator to easily separate device pools and permissions across different research projects or courses.
User groups and enrollment codes can both be made to expire after a certain duration, such as at the end of a research collaboration or academic term.

\begin{figure*}[!t]
	\centering
	\includegraphics[width=\textwidth]{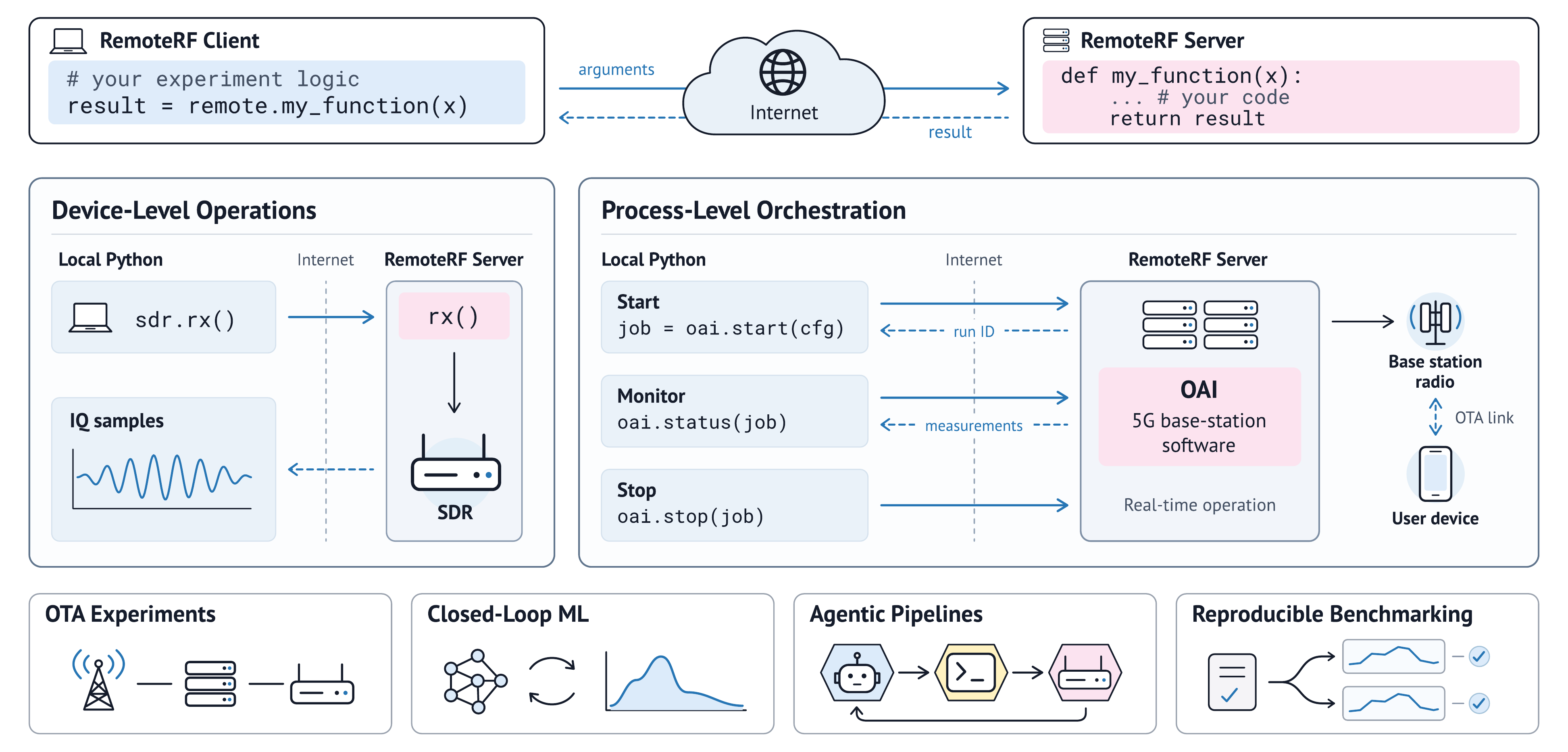}
	\caption{\rev{Example research workflows on RemoteRF: server-side Python wrappers invoked as networked RPCs support both device-level operations and process-level orchestration such as OpenAirInterface (OAI), enabling over-the-air experiments, closed-loop ML, agentic pipelines, and reproducible benchmarking.}}
	\label{fig:research}
\end{figure*}

\subsection*{\rev{Device Support}}
\rev{RemoteRF natively supports a wide range of popular \sdrs---such as USRPs, ADALM-PLUTO, Pluto+, HackRF, RTL-SDR, and KrakenSDR---as well as peripherals like radars, cameras, and GPS receivers.}
Beyond these, administrators can onboard custom devices that are not natively supported by writing simple Python wrappers that define their functionality and properties.
These custom wrappers are written on the RemoteRF server and automatically pulled by the RemoteRF client in the background, requiring no software updates or changes on the end user's side.
This makes it possible for administrators to expand their own RemoteRF deployment without any intervention from the RemoteRF development team.
\rev{As illustrated in \figref{fig:research}, the same wrapper mechanism also allows administrators to create high-level server-side \glspl{rpc} that execute multiple functions or entire scripts remotely from a single user command, such as running a multi-step calibration routine or launching a full-stack 5G network via \gls{oai}.}

\subsection*{\rev{Software Lockouts and Spectrum Management}}
Because device control happens server-side, software lockouts can limit \sdr operation---for example, capping transmit power and restricting carrier frequency to ensure regulatory compliance. Hardware filters and attenuators can also be installed on the \sdrs as an additional safeguard.
\rev{Additionally, users can query the operating frequencies of other active devices on a deployment directly from Python, allowing them to coordinate spectrum use and avoid co-channel interference during concurrent experiments.}

\section{\rev{Local and Global RemoteRF Deployments}}

\subsection*{\rev{Local Deployments}}
In its simplest configuration, a RemoteRF server operates entirely on a local network or behind an institution's \vpn, keeping all user accounts, reservations, and device access within its private network. A lab or department can set up a local deployment in minutes for under \$1,000~USD with just a few \sdrs, and later expand it as necessary into a larger testbed spanning rooms, buildings, and rooftops using lightweight hosts that require only power and a network connection, as illustrated in \figref{fig:edgehosts}. From a user's perspective, distributed host devices appear and function identically to hardware connected directly to the server.

\rev{The flagship deployment at UCLA, launched in early 2025 and shown in \figref{fig:edgehosts}, illustrates how a testbed can grow over time. It began with five ADALM-PLUTO \sdrs and has since expanded to over 30 devices housed in a central server rack and in weatherproof host enclosures across campus. Its device pool now includes ADALM-PLUTOs, USRPs, RTL-SDRs, and a KrakenSDR alongside radar, 3D camera, GPS, and Wi-Fi/Bluetooth peripherals. To support progressive testing from controlled channels to the real world, these radios are deployed in several physical configurations: single-device and two-device coaxial loopbacks, multi-tap coaxial loopbacks that introduce controlled multipath, and indoor and outdoor \gls{ota} links. Devices are connected to the server via controllable USB hubs and \gls{poe} adapters to automatically power down idle radios and allow users to remotely power-cycle hardware if needed.}\vspace{0pt plus 2.5ex}\break
\subsection*{\rev{Global Deployments}}
\begin{figure}[!t]
	\centering
	\includegraphics[width=\linewidth]{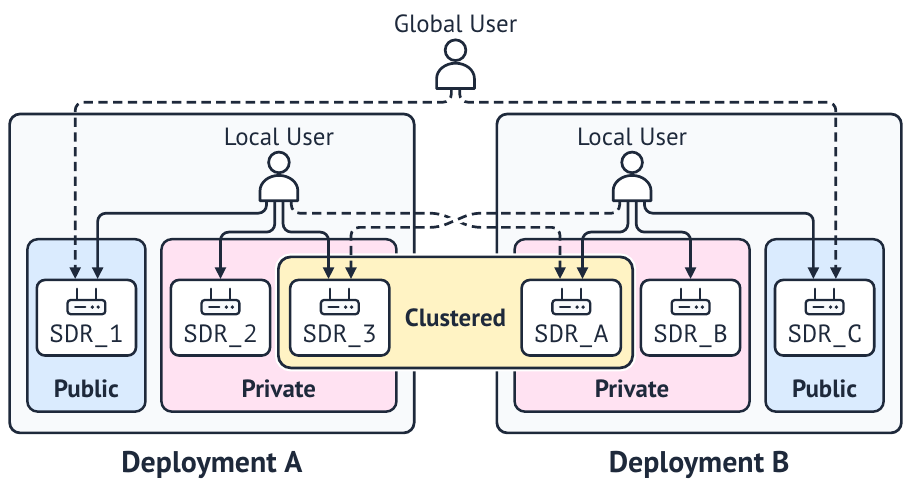}
	\caption{\rev{Device sharing modes on RemoteRF Global: private access, multi-institution clustering across partner deployments, and public access for verified users worldwide.}}
	\label{fig:deployments}
\end{figure}

\rev{To make a deployment accessible directly over the internet---without requiring users to be on a local network or campus \vpn---administrators can optionally connect their server to the \textit{RemoteRF Global} network, as UCLA has done with its own deployment. Connecting a deployment to RemoteRF Global does not automatically make its hardware public or relinquish control: the local server remains the sole authority over its devices and reservation schedule, and all devices stay private by default. Using the same user group and enrollment code framework as a local deployment, administrators can configure three sharing modes over the internet, as shown in \figref{fig:deployments}:
\begin{itemize}
    \item \textbf{Private Access:} Administrators can keep their deployment private to their own students, researchers, or external collaborators via enrollment codes, giving them convenient access over the internet from anywhere without needing a campus \vpn.
    \item \textbf{Multi-Institution Clustering:} Two or more deployments connected to RemoteRF Global can link together to form a private cluster across partner institutions. This allows their users to share access to select devices and run experiments across multiple deployments without exposing their hardware publicly.
    \item \textbf{Public Access:} Deployments can also make a subset of their devices public to any verified user on RemoteRF Global, enabling students and researchers anywhere in the world to run experiments on real hardware.
\end{itemize}}

\subsection*{\rev{Latency and Scalability}}
\begin{figure}[!t]
	\centering
	\includegraphics[width=\columnwidth]{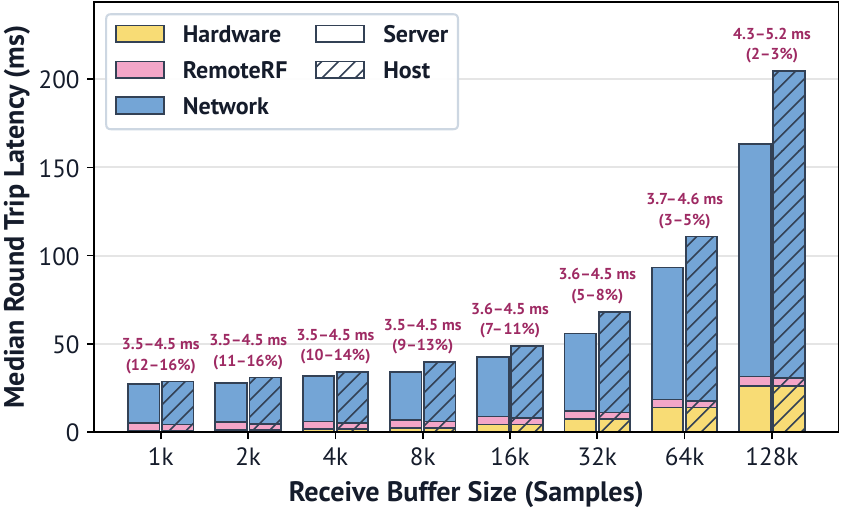}
	\caption{\rev{Median end-to-end latency of fetching received samples via \texttt{sdr.rx()} on an ADALM-PLUTO SDR connected directly to the server and via a host, decomposed into hardware, RemoteRF, and network latency. Numbers above the bars indicate the portion attributed to the RemoteRF platform itself.}}
	\label{fig:latency}
\end{figure}

\rev{To quantify the latency introduced by the platform, we benchmarked the median end-to-end latency of fetching received samples via \texttt{sdr.rx()} on an ADALM-PLUTO \sdr connected directly to the server and via a Raspberry Pi host, accessed by an off-campus public user over RemoteRF Global. \figref{fig:latency} decomposes the total time across buffer sizes from 1k to 128k samples into hardware capture, RemoteRF software overhead, and network transfer. Across all buffer sizes, the overhead added by RemoteRF itself remains nearly constant at just 3.5--5.2\,ms, dropping from 12--16\% of the total latency at 1k samples to 2--3\% at 128k samples as network transport and hardware capture dominate. Notably, RemoteRF overhead on the host happens to be slightly lower in this case due to network punch-through---which routes traffic directly between the client and host---whereas its network latency is higher because the Raspberry Pi is on a Wi-Fi connection rather than the server's wired \gls{lan}.}

\rev{When transmitting and receiving waveforms from a user's computer, each burst is still transmitted or captured by the \sdr at its full hardware sampling rate using onboard buffers, while network speed and latency govern how quickly buffers are transferred between the radio and client across calls.
To reduce this transfer latency, RemoteRF Global uses peer-to-peer punch-through where possible to avoid an intermediate relay hop.
For real-time applications or experiments with strict latency constraints, operations can instead be executed directly on the server or host via custom wrappers rather than streaming samples to the client.
Meanwhile, since signal processing typically happens on client machines, a single modest server can comfortably support dozens of active devices and users. In tests on the UCLA server, end-to-end latency remained nearly constant even with 16 concurrent users.}

\section{\rev{Wireless Research with RemoteRF}}

\subsection*{\rev{Over-the-Air Experimentation}}
As illustrated before in \figref{fig:research}, \gls{ota} experimentation with RemoteRF allows researchers to capture both real-world propagation effects and hardware nonidealities such as phase noise, IQ imbalance, and power amplifier nonlinearity. Researchers can first validate algorithms on cabled or multi-tap loopback devices before moving to \gls{ota} links simply by changing the device handle in their Python script. A single script can also reserve and coordinate multiple \sdrs alongside multi-modal peripherals for joint communication and sensing experiments. At UCLA, researchers are using RemoteRF in this way across several projects, including evaluating phased-array beamforming methods for interference cancellation and \gls{isac}, and studying communications-constrained robotics by mounting a RemoteRF host directly onto a mobile robot.

\begin{figure*}[!t]
    \centering
    \subfloat[\rev{Campus-scale SNR heatmap from a rooftop host.}\label{fig:research_examples_a}]{%
        \includegraphics[height=2.35in]{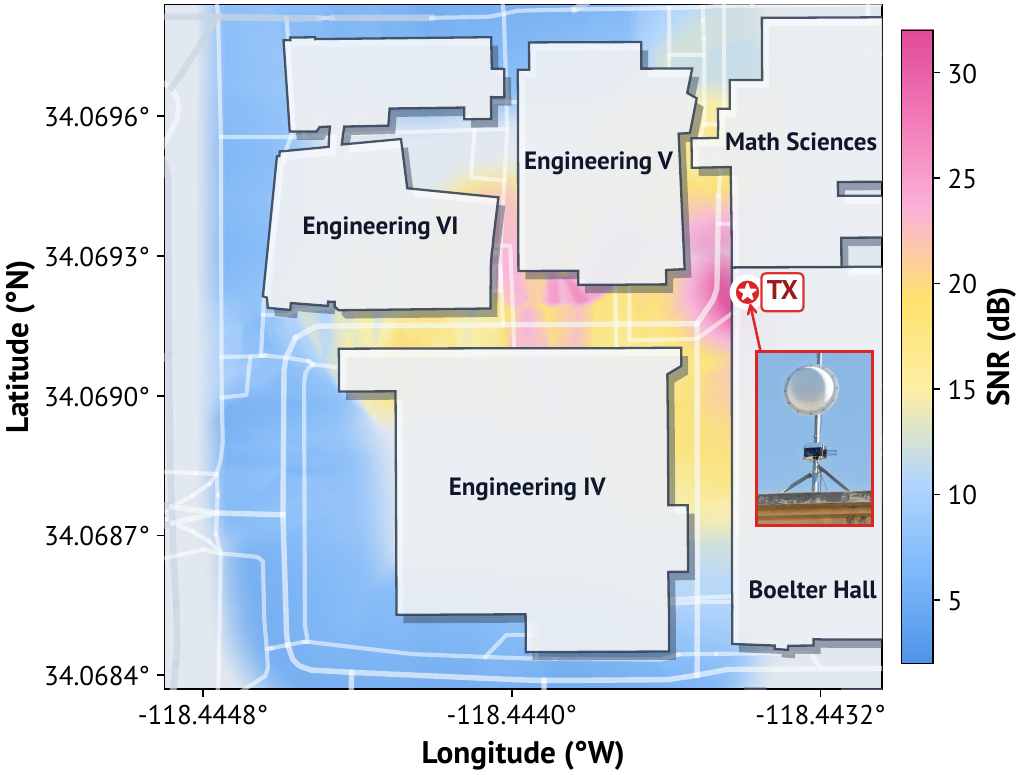}%
    }\hspace{0.18in}%
    \subfloat[\rev{Over-the-air BER under co-channel burst interference.}\label{fig:research_examples_b}]{%
        \includegraphics[height=2.35in]{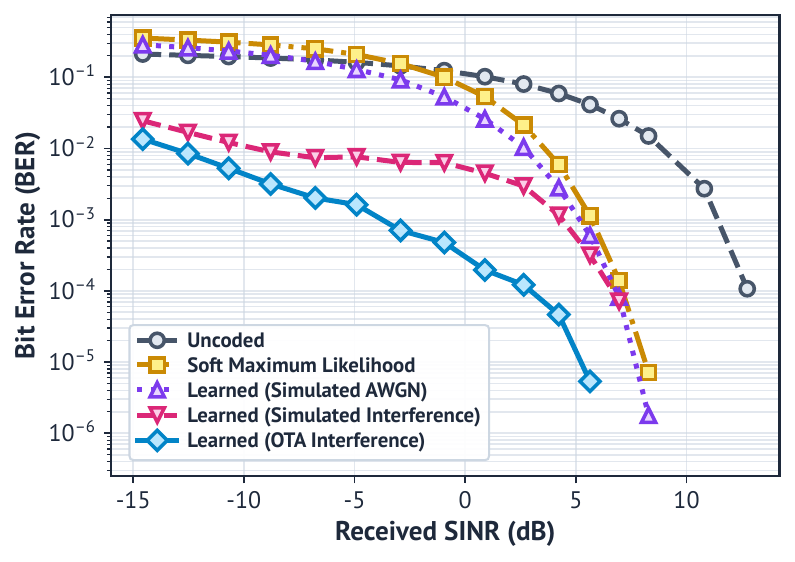}%
    }
    \caption{\rev{Example research experiments on the UCLA deployment: (a)~received SNR measurements collected across campus from a rooftop host transmitting atop Boelter Hall, and (b)~over-the-air BER versus SINR under burst interference with classical and learned decoding.}}
    \label{fig:research_examples}
\end{figure*}

\subsection*{\rev{Site-Specific Testbeds and Digital Twins}}
\rev{Using RemoteRF hosts to distribute \sdrs across buildings or an entire campus allows researchers to study site-specific wireless propagation and system performance in a given physical environment. Pairing these radios with multi-modal peripherals such as GPS receivers and cameras also makes it straightforward to collect measurements to calibrate, augment, and validate ray-tracing models and digital twins~\cite{alkhateeb_2023_twins}. As shown in \figref{fig:research_examples_a}, for example, we collected received power measurements across the walkways surrounding UCLA's engineering buildings from a rooftop RemoteRF host transmitting atop Boelter Hall, capturing site-specific shadowing and urban canyon propagation effects. Just as importantly, whereas experimental wireless researchers have traditionally evaluated algorithms on bespoke local setups that are difficult for others to perfectly reproduce~\cite{marti_2026_sionna_bench}, sharing a deployment through RemoteRF Global lets researchers across institutions benchmark algorithms over the air at the exact same physical site, using the same hardware and propagation environment, much like open datasets have enabled common benchmarks in \ml/\ai.}

\subsection*{\rev{Full-Stack 5G/6G}}
\rev{Experimental wireless research is often divided between physical-layer prototyping---where researchers transmit and receive signals using custom techniques---and higher-layer systems research on full-stack testbeds such as \gls{oai}. RemoteRF bridges this gap by allowing the same shared hardware pool to be flexibly used for both: for example, a single physical USRP can be used to stream raw IQ samples in Python or launched as an \gls{oai} base station or user equipment node. Researchers can thus deploy standards-compliant 5G and emerging 6G networks and evaluate new techniques in terms of full-system, end-to-end performance without maintaining a separate testbed. Researchers can use these \gls{oai} links to extract physical-layer measurements such as \gls{csi} alongside protocol-level telemetry from live 5G/6G signals, enabling research on initial access, beam management, scheduling, and cross-layer design.}

\subsection*{\rev{Data Collection for ML/AI Methods}}
Because \ml/\ai models trained solely on synthetic data often degrade under real hardware impairments and propagation effects, training and validating models that generalize in practice requires real-world \gls{ota} measurements. RemoteRF simplifies collecting such datasets at scale from a single Python script that systematically iterates across multiple transmitters, receivers, carrier frequencies, and physical locations. Furthermore, since received signals are returned directly to the user's computer, live \gls{ota} captures can be piped straight into local GPU workflows for dataset curation, model training, or online adaptation.

\rev{To illustrate the benefit of live \gls{ota} adaptation beyond offline simulation, \figref{fig:research_examples_b} presents results from an experiment conducted at 915\,MHz on the UCLA deployment. One ADALM-PLUTO transmits interleaved rate-$1/2$ $(16,8)$ block-coded QPSK frames to a receiver while a nearby transmitter creates bursts of narrowband interference with random duration and duty cycle. Under strong interference, classical soft maximum-likelihood decoding performs worse than uncoded QPSK because corrupted bit metrics from a single affected symbol overwhelm clean symbols in the same codeword. Even when trained in simulation under the same interference model, a deep learning-based receiver lacks the interferer's actual frequency offset, its propagation channel, and the radios' front-end response; consequently, it learns to simply ignore interfered symbols and rely on parity bits, plateauing whenever too many symbols in a codeword are affected. Fine-tuning the model on just a few seconds of \gls{ota} captures lets it adapt to these hardware and channel characteristics, allowing it to suppress the interference and recover bits correctly even during strong bursts.}

\subsection*{\rev{AI-Driven Autonomous Experimentation}}
\rev{Autonomous \ai agents are increasingly used to write signal processing code, run simulations, and iterate on algorithms, opening the door to automating experimental wireless research~\cite{bjornson_2026_what_remains_human}. Putting agents in the loop with real radios, however, requires programmatic ways to allocate and control hardware alongside strict safeguards against unsafe or out-of-band transmissions. RemoteRF provides both through a built-in \gls{mcp} server and installable agent skills that allow \ai agents to discover available devices, make reservations, and execute Python scripts across deployments, while reservation limits and server-side lockouts keep operation within safe and regulatory bounds. An agent can thus autonomously write Python code, transmit and receive over the air, evaluate metrics such as bit error rate or error vector magnitude, and iteratively refine its algorithms based on real-world measurements. In fact, the \gls{ota} experiment in \figref{fig:research_examples_b} was executed end-to-end by \ai agents in precisely this way using RemoteRF's \gls{mcp} interface.}

\section{\rev{Engineering Education with RemoteRF}}
\rev{In most wireless engineering courses, students work primarily through mathematical problem sets and software simulations, rarely getting the chance to test their designs on real hardware. Rather than replacing these traditional tools, RemoteRF complements them: students can first verify their designs in simulation before moving to actual radios, where they are forced to confront and address real-world effects such as frequency and timing offsets, hardware nonlinearities, and \gls{ota} interference. Beyond building practical debugging skills that are especially valuable in the era of \ai~\cite{bjornson_2026_what_remains_human}, getting an \gls{ota} link working is also exciting and fulfilling for many students in a way that simulations alone rarely match.}

\rev{To incorporate \sdrs into a course, an instructor can either deploy a local RemoteRF server on campus with minimal cost and setup or use publicly shared devices on the RemoteRF Global network without purchasing any hardware at all.} Through RemoteRF, students in virtually any sized class can share access to a limited number of \sdrs and complete lab exercises at any time from anywhere with an internet connection. The flexibility afforded by this anytime--anywhere access can be particularly valuable to non-traditional students, such as those who have caregiving responsibilities, commute to campus, or take classes online.

\subsection*{Undergraduate Courses}
In January 2025, RemoteRF made its debut at UCLA in an undergraduate course on communications and has been used each academic quarter since then in multiple courses.
In that first course, 60 undergraduate students used RemoteRF to complete lab exercises where they implemented digital modulation, pulse shaping, matched filtering, symbol detection, channel estimation, and equalization.
Sharing just the initial five \sdrs on the UCLA deployment, all 60 students were able to complete their biweekly labs on time, highlighting the massive cost savings compared to purchasing \sdrs for each student or holding in-person lab sessions led by teaching assistants.
Two of the five \sdrs were configured with a cabled loopback connection for initial debugging, while the other three \sdrs had antennas connected to their transmit and receive ports, with a band-pass filter inserted at the transmitter to ensure students could only operate within the 915 MHz \gls{ism} frequency band.
Following this initial offering, additional \sdrs and channel configurations were added to the server to increase availability for larger classes and to support more diverse lab exercises.

\subsection*{Graduate Courses}
Beyond undergraduate instruction, RemoteRF has also been used in graduate-level courses at UCLA. In a graduate course on digital communications, for example, students developed timing and frequency synchronization techniques to successfully transmit from one \sdr to another over the air.
Students were also tasked with implementing multi-tap channel equalization and \gls{ofdm} to combat frequency selectivity using a multi-tap loopback channel built from coaxial cables of different lengths.
This course is co-listed with UCLA's online master's degree program, and given RemoteRF can be accessed over the internet, these remote learning students were able to complete all of the same lab exercises as on-campus students.
In another graduate course on wireless communications, students used RemoteRF to complete quarter-long research projects where they implemented \gls{otfs} modulation, 5G initial access mechanisms, and \ml-based channel coding techniques.

\subsection*{\rev{Usage Statistics and Student Feedback}}
\rev{To date, the UCLA deployment has logged over 8,000 reservations by more than 250 researchers and students, totaling over 4,000 hours of use. Usage logs show how users take advantage of the platform's continuous availability and automated reservation manager. On a typical weekday, 62\% of usage takes place outside normal business hours of 9\,AM to 5\,PM, with heavy activity on weekends as well. Furthermore, reservation lead times show that users can reliably access hardware when they need it: nearly 60\% of reservations are made for the ongoing time window (i.e., instant access), and another 24\% start within 30 minutes. This confirms that a relatively small pool of devices can support a much larger user base without encountering serious scheduling conflicts.}

Anonymous surveys of 50 students before and after using RemoteRF also reflect its value in the classroom: while 88\% of students expressed strong interest in using \sdrs beforehand, 86\% had never used an \sdr prior to RemoteRF. After using the platform, 92\% reported that it improved their understanding of course concepts, 86\% said it helped reinforce prerequisite signal processing material, and 83\% said it increased their interest in wireless communications and RF engineering.

\section{\rev{Conclusion and Outlook}}
By allowing users to reserve and control remote \sdrs directly from Python on their own computers, RemoteRF lowers many of the financial and logistical barriers to incorporating real hardware into wireless research and education. RemoteRF's open-source software and documentation are freely available at \texttt{remoterf.net}, enabling anyone to deploy their own instance in minutes \rev{or access shared hardware on RemoteRF Global today}.

\rev{While RemoteRF already supports a broad range of research and educational workflows, several technical extensions are planned for future releases. A top priority is enabling tight time, frequency, and phase synchronization across distributed \sdrs---through GPS-disciplined oscillators and \gls{ota} calibration routines---to support coherent multi-node experiments such as distributed MIMO and multistatic sensing. We are also expanding native support to include phased arrays, channel emulators, robotic/motorized platforms, and additional \sdrs and multi-modal sensors, as well as developing standardized benchmark suites for evaluating new techniques under common hardware and propagation conditions. Our hope is that researchers, instructors, and students worldwide will use RemoteRF in their own labs and classrooms, and join us in growing a shared global network of wireless testbeds.}

\section*{Acknowledgments}

This work has been supported by the Educational Innovation Grants program and the Internet Research Initiative at UCLA.
We would also like to acknowledge Hanju Yoo and Dominic Olson for their contributions to RemoteRF.

\bibliographystyle{bibtex/IEEEtran}
{\small \bibliography{refs}}

\end{document}